\documentclass[preprint,aps,pra,12pt]{revtex4-1}
\usepackage{amsmath}
\usepackage{graphicx}
\usepackage[caption=false]{subfig}

\begin{document}

\title{Temporal laser pulse shape effects in nonlinear Thomson scattering}

\author{V.~Yu.~Kharin}
\email{v.kharin@gsi.de}
\affiliation{Helmholtz-Institut Jena, Fr\"obelstieg 3, 07743 Jena, Germany}
\author{D.~Seipt}
\affiliation{Friedrich-Schiller-Universit\"at, Theoretisch Physikalisches
Institut, 07743, Jena, Germany}
\author{S.~G.~Rykovanov}
\affiliation{Helmholtz-Institut Jena, Fr\"obelstieg 3, 07743 Jena, Germany}

\date{\today}

\begin{abstract}

    The influence of the laser pulse temporal shape on the Nonlinear Thomson
    Scattering on-axis photon spectrum is analyzed in detail. Using the
    classical description, analytical expressions for the temporal and spectral
    structure of the scattered radiation are obtained for the case of symmetric
    laser pulse shapes. The possibility of reconstructing the incident laser
    pulse from the scattered spectrum averaged over interference fringes in the
    case of high peak intensity and symmetric laser pulse shape is discussed.

\end{abstract}


\maketitle
\section{Introduction}

During the last decades scattering of light on high-energy electron beams has
become an indispensable tool for generating tunable wide range X- and
$\gamma$-radiation\cite{Esarey:PRE1993, Leemans1996, Albert2010,Karagodsky2011,
Rykovanov2013,khrennikov2015tunable,sarri2014ultrahigh,powers2014quasi}. These
Thomson scattering (TS) sources have found applications in
spectroscopy\cite{Nedorezov2004, Thirolf2012, Tesileanu2013, Johnson2011},
medicine\cite{Weeks1997c}, ultrafast radiography\cite{Tommasini2011,
Toyokawa2002}, and nuclear nonproliferation\cite{Bertozzi2008, Quiter2011,
Geddes2015}. The conversion of laser photons to $\gamma$-radiation occurs as the
Doppler upshift of the laser frequency while scattering on electrons with high
gamma-factor. Since the TS cross-section is relatively small, for the generation
of bright X- and $\gamma$-ray beams one benefits from employing high-intensity
incident laser pulses. However, the longitudinal drift of the electron in the
strong wave -- caused by the $\mathbf{v}\times \mathbf{B}$ force -- results in
significant shift of the radiation frequency.  This gives rise to the broadening
of the spectral components of the scattered light and a specific interference
structure inside each spectral harmonic\cite{Hartemann:PRE1996, Brau2004,
    Krafft:PRL2004, Mackenroth2011, Seipt:LasPhys2013, Seipt:PRA2011,
Rykovanov:arxiv2014}.  A number of recent works were devoted to the studies of
such a ponderomotive broadening, and its compensation by the usage of laser
chirping techniques\cite{Ghebregziabher:PRSTAB2013, Terzic2014, Seipt:PRA2015,
Rykovanov:arxiv2014}. However, a quantitative analytical description of the
nonlinear TS still remains an open question. 

In this paper, the temporal and spectral structure of the on-axis scattered
radiation for nonlinear TS is found analytically within the assumption of a
symmetric laser pulse temporal shape for both linear and circular laser pulse
polarization. It is demonstrated that the shape of the envelope of the scattered
spectrum permits the reconstruction of the temporal structure of the strong
incident laser pulse. This could be useful for intense laser-matter
interaction experiments.

The paper is organized as follows. In Sec. \ref{sec:analytical} the analytical
description of the nonlinear Thomson back-scattering is provided within
classical electrodynamics. That means, equations of motion for electron are
solved, and the scattered radiation is calculated using Lienard-Wiechert
potentials. The detector time consideration described in this section provides
also an efficient way of calculating the scattered spectrum numerically.  This
is covered in Sect.~\ref{sec:numerical}, where we provide the details of a fast
numerical routine for evaluation of the back-scattered spectrum.
Subsect.~\ref{subsec:results} contains the scattered spectra for various laser
pulse shapes. The details of the scattered pulse formation in time domain allow
us to make qualitative statements about the spectrum. The shape of the
back-scattered spectrum can be evaluated analytically with good accuracy.  Using
this approach, one sees that the shapes of the scattered light spectra at
relatively high intensity differ a lot, even when the envelope profile
difference is not very well pronounced. It is shown, that in case of the
symmetric pulse envelope and high peak intensity one can reconstruct the
incident pulse envelope profile without resolving the interference structure of
the nonlinear TS.  In the conclusions (Sect. \ref{sec:concl}) the results are
summed up, and the possible influence of the effects not included in the model
is discussed. An Appendix contains derivation of the scattered spectral
intensity in case of linearly polarized pulse and shapes of spectral intensity
profiles for the pulses used in the article

\section{Analytical description}\label{sec:analytical}

Throughout the paper the classical description of TS process is used. That is,
the classical equations of motion of the electron are solved, and the scattered
field is calculated using Lienard-Wiechert\cite{lienard1898champ,
wiechert1901elektrodynamische} potentials. Within this framework the electron
recoil is assumed negligible.  For this case the structure of integrals defining
spectral intensity remains similar to the one in the first order perturbation
theory in quantum electrodynamical framework (cf.  \cite{Narozhnyi:JETP1996,
Seipt:LasPhys2013} and \cite{Rykovanov:arxiv2014}).  Despite the simplicity of
the underlying model, even the scattering of quasi-monochromatic plane wave
radiation demonstrates a rich physics, as will be shown further.

The influence of spin effects on the scattered spectra and comparison between
classical and QED solutions can be found in \cite{Krajewska:LasPB2013,
Krajewska:PRA2014}. Both the classical and quantum mechanical approaches
mentioned above use the approximation that the strong incident laser pulse is
described as a purely classical. Description of the strong-field dynamics in
quantized external field is another complicated and interesting problem. 

\subsection{General setup}
\label{subsec:general}

Throughout the paper the natural units for the relativistic kinematics in the
laser field are adopted. In these dimensionless units
\begin{align*}
    t = \omega_L\hat{t},\quad 
    x = k_L \hat{x},\quad
    y = k_L \hat{y},\quad
    z = k_L \hat{z},\quad
    &A = \frac{e\hat{A}}{m_ec^2},
\end{align*}
where the ``hatted'' values are given in Gaussian CGS units, $t$, $x$, $y$ and
$z$ are the natural time and coordinates, $A$ is the normalized vector
potential, $e$ and $m_e$ are the electron absolute charge and mass,
respectively, $c$ is the speed of light in vacuum, $\omega_L$ is the central
laser pulse frequency, and $k_L = \omega_L/c$.

Since one can always apply the Lorentz transformation to the electron frame of
reference, we consider the electron initially at rest keeping in mind that for
real situation the calculated spectra should be Lorentz transformed back.  It
should be mentioned, that in this frame of reference the scattered light will
not be Doppler up-shifted because the initial gamma-factor of the electron is 1.
We also constrain ourselves to the case of the on-axis scattering. The equations
of motion for the charged particle in the plane wave field can be solved
analytically\cite{LandauVol2}. One can choose the gauge where the vector
potential $\mathbf{A}^L$ of the laser is transverse (i.e., it has only $x$ and
$y$ components), and the wave is propagating in $z$ direction:
\begin{equation}
    \mathbf{A}^L=\mathbf{A}^L(t-z).
\end{equation}
Superscript $L$ stands for ``laser'' (not to confuse the laser field with the
vector potential $\mathbf{A}$ describing the scattered field). We use the
notation $\varphi = t-z$ for the laser pulse phase, $\zeta = t+z$.  Moreover,
we require that the laser vector potential vanishes asymptotically,
$\mathbf{A}^L\rightarrow 0$ at $\varphi\rightarrow\pm\infty$.  The analytic
solution of the equations of motion for components of electron four-velocity
$u$ reads
\begin{equation}
    \label{eq:guz}
    \gamma - u_z = 1,
\end{equation}
\begin{equation}\label{eq:ux}
    u_{x,y} = A_{x,y}^L, 
\end{equation}
\begin{equation}
    \gamma+u_z = 1 + |\mathbf{A}^L|^2\,,
    \label{eq:guzp}
\end{equation}
\begin{equation}
    \label{eq:zeta}
\zeta = \varphi + \int\limits_{0}^{\varphi}|\mathbf{A}^L(\xi)|^2d\xi,
\end{equation}
where $\gamma$ is the Lorentz factor of the electron.  As there is no
significant difference between the $x$ and $y$ components of the scattered
radiation, one can work with only one component ($x$, for example). The
Lienard-Wiechert formula for the vector potential of the back-scattered light
that arrives at the distance $R$ from the initial electron position at time $t$
can be written as follows
\begin{equation}
    A_x(t) =\frac{1}{R-z}\left.\frac{u_x}{\gamma+u_z}\right|_{t_{ret}}\mathrm{,}
\end{equation}
where $t_{ret}$ is the retarded time when the light wave was emitted by the
electron. As the far-field distribution of the radiation is the subject of
interest, one can omit the change in the electron position $z$ in the denominator,
since it contributes only to terms decreasing faster than $1/R$ with $|z|\ll
R$. Hence, the following relation between the retarded time $t_{ret}$ and the
detector time $t$ can be written:
\begin{equation}
    t=R+z(t_{ret})+t_{ret}.
\end{equation}
If one now tunes the detector clock to the delay, $t\mapsto t+R$, one can
immediately see that the detector time is exactly equal to the variable $\zeta$
defined above. Since the explicit solution of the electron equation of motion, Eqs.
(\ref{eq:guz}-\ref{eq:zeta}), is written in terms of the pulse phase
$\varphi$, and there is the expression for $\zeta$, it is convenient to
eliminate the $t_{ret}$ dependence, and get the expression for the vector
potential of the emitted radiation as
\begin{equation}
    \label{eq:At}
    \mathbf{A}(\zeta)=\frac{1}{R}\frac{\mathbf{A}^L(\varphi(\zeta))}{1+|\mathbf{A}^L(\varphi(\zeta))|^2}.
\end{equation}
Here the variable $\zeta$ is precisely the detector time shifted by the constant
value of $R$, and the dependence $\varphi(\zeta)$ is given implicitly by
(\ref{eq:zeta}).  Taking the Fourier transform of (\ref{eq:At}) with respect to
detector time and changing the variables of integration from $\zeta$ to
$\varphi$ yields the following simple expression for the spectrum of
$\mathbf{A}$:
\begin{equation}
    \mathbf{A}(\omega)= \frac{1}{R}\int\limits_{-\infty}^{+\infty}\mathbf{A}^L(\varphi)e^{i\omega\zeta}d\varphi.
\end{equation}

Both the cases of circularly and linearly polarized intense laser light with
slowly varying amplitude can be considered to provide the analytical description
of the back-scattered radiation spectrum. 

\subsection{Circular polarization}
\label{subsec:circular}
Let us start with the case of circular laser polarization, since, as we shall
see soon, this simplifies the analytical
expressions. We take the pulse in the following form
\begin{align}
    &A^L_x(\varphi)=a(\varphi)\cos\varphi,\\
           &A^L_y(\varphi)=a(\varphi)\sin\varphi\mathrm{,}
\end{align}
where $a(\varphi)$ is a slowly varying amplitude. 
With this laser vector potential Eq. (\ref{eq:zeta}) can be written as
\begin{equation}
    \label{eq:zeta_phi}
    \zeta = \varphi + \int\limits_{0}^{\varphi}a(\xi)^2d\xi\,,
\end{equation}
where the non-linear term depends only on the envelope of the vector potential
$a(\varphi)$, and there are no oscillations (with the frequency of the order of
unity) in it. Physically, this leads to the absence of high order harmonics in
the on-axis back-scattered radiation\cite{HeinzlPRA2010}.  One can consider only one
component of the scattered field, since the reasoning for the other component
(and the spectra) will be the same. We rewrite the scattered vector potential in
terms of $a(\varphi)$ as
\begin{equation}
    \label{eq:At_circ}
    A_x(\zeta)=\frac{1}{R}\frac{a(\varphi(\zeta))}{1+a(\varphi(\zeta))^2}\cos\varphi(\zeta),
\end{equation}
From now on we will refer to $a_0$ as the peak value of $a(\varphi)$. The
argument of the instant amplitude dependence $a(\varphi)$ will be omitted when
it cannot lead to confusion. From (\ref{eq:At_circ}) one can clearly see that if
$a_0\ll 1$ the scattered pulse reproduces the incident one,
$A_x(\zeta)=A_x^L(\zeta)/R$, because $\zeta\approx\varphi$. When $a_0$
increases, two effects contributing to the scattered pulse modification can be
noticed. The first one is due to the nontrivial relation between laser phase and
detector time $\zeta(\varphi)$.  Physically, it is an ordinary time-dependent
Doppler shift, and, hence, it non-linearly ``stretches'' the pulse of emitted
radiation. The second effect is an amplitude damping due to the denominator in
(\ref{eq:At_circ}). To make things more clear we discuss these two effects
separately before proceeding to analyze the spectral properties of the
back-scattered radiation.

\subsubsection{Detector time stretching}
\label{ssubsec:doppler}

\begin{figure}
    \subfloat[Circular polarization]{
        \includegraphics[width=.5\linewidth]{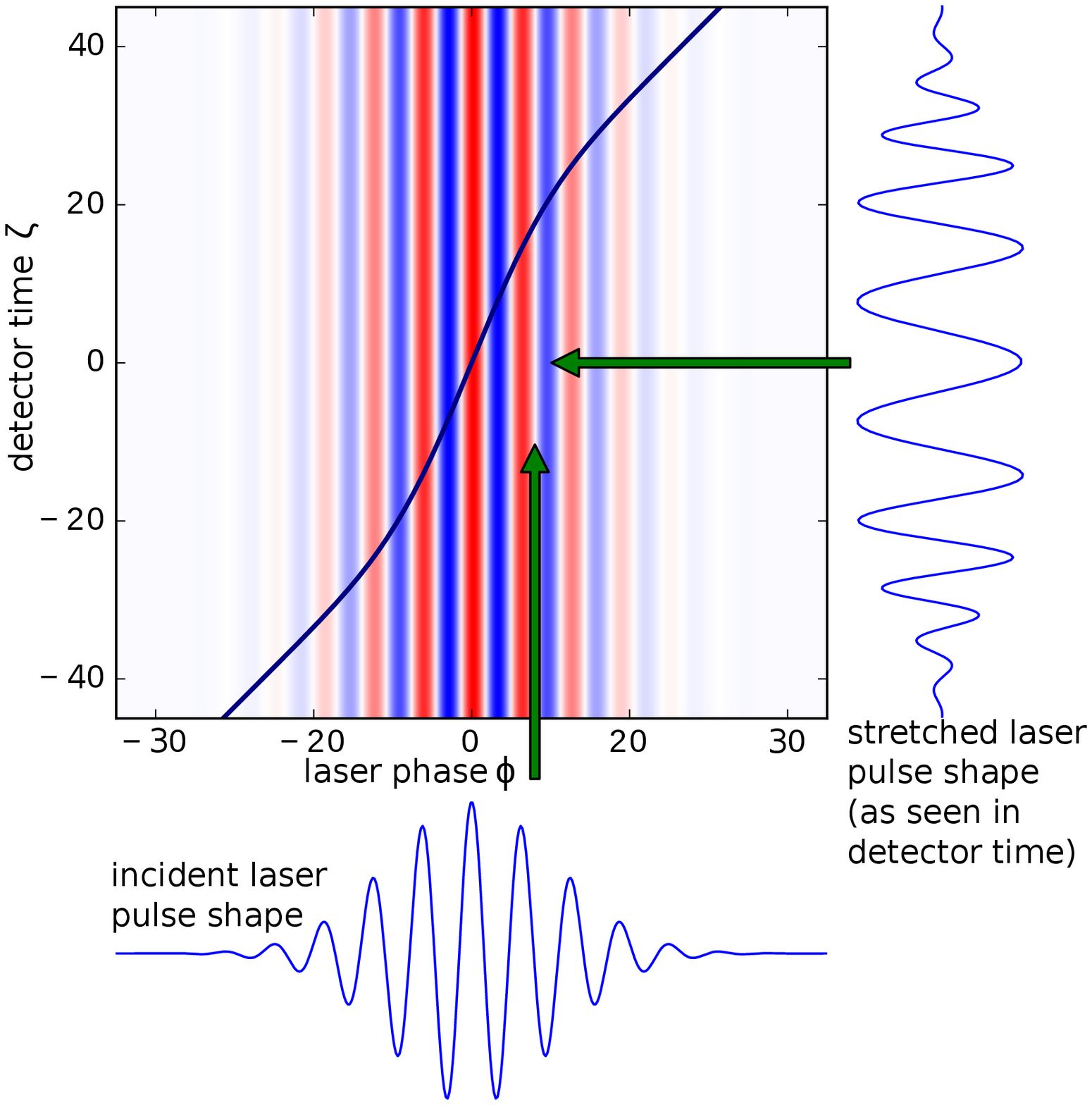}
        \label{fig:dilation0}
    }
    \subfloat[Linear polarization]{
        \includegraphics[width=.5\linewidth]{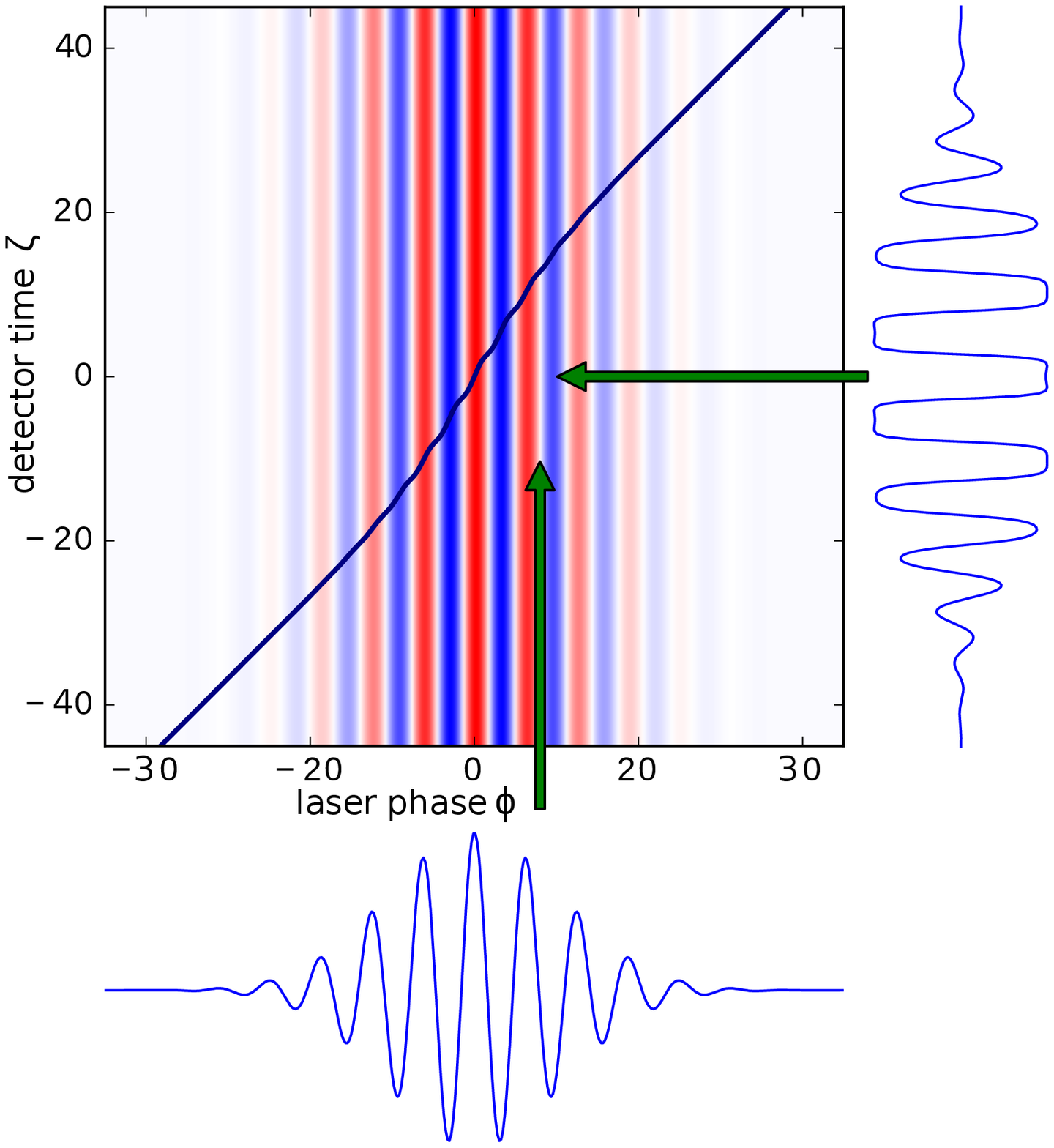}
        \label{fig:dilation1}
    }
    \centering
    \caption{
    \label{fig:dilation}
        $A^L(\varphi)$ is the color-coded surface. Dark line presents the $\zeta(\varphi)$
        dependence (\ref{eq:zeta_phi}). Projection of the surface section by $\zeta(\varphi)$
    on $\zeta$ axis yields $A^L(\zeta)$.}
\end{figure}

From the physical point of view the term $\int\limits_{0}^{\varphi}a(\xi)^2d\xi$
in (\ref{eq:zeta_phi}) means that the electron rides the laser wave, so, during
the interaction one observes a Doppler red-shifted signal on the detector.
Hence, looking at the detected signal, one can say, that the initial pulse is
``stretched'' during the interaction. Fig.~\ref{fig:dilation} illustrates this
process. Here color-coded surface reproduces $A^L(\varphi)$ dependence with
trivial dependence on $\zeta$. The intersection of this surface with
$\zeta(\varphi)$ provides ``stretched'' pulse when projected on $(A,\zeta)$
plane. The oscillations on $\zeta(\varphi)$ curve in case of linearly polarized
pulse are discussed below.

Formally, the ponderomotive term $\int^\varphi a^2(\xi)d\xi$ is nothing else,
but smooth non-decreasing function which asymptotically equals to zero at
$\varphi=0$ and some constant at $\varphi\rightarrow\pm\infty$. It can be
thought of as a smooth approximation for the step function. For example, for the
Gaussian pulse it will be equal to an error function plus a constant.  If the
amplitude is slow-varying, one can see the obvious consequence: the scattered
pulse will have a chirp due to the factor of $\cos\varphi(\zeta)$ in the
scattered vector-potential (that means, that the scattered frequency will vary
with time), which will follow the shape of the incident pulse intensity. 

\subsubsection{Amplitude modulation}
\label{ssubsec:amplitude}

Here, we shall focus on the envelope of the scattered pulse without dealing with
the pulse stretching. As has been mentioned before, for $a_0^2\ll1$ the envelope
deformation can be neglected.  When the amplitude grows, one must take the
denominator of (\ref{eq:At_circ}) into account. The amplitude of the field
undergoes the following transformation
\begin{equation}
    f:x\longmapsto\frac{x}{1+x^2}.
\end{equation}
It is easy to see that it has a maximum at $x=1$. This means the maximum
amplitude of the scattered vector potential will be achieved when the
instantaneous amplitude of the pulse is $a=1$. The amplitude of light scattered
by those parts of the laser pulse with an amplitude exceeding this value will
be suppressed. It is also notable, that if the incoming pulse has a single peak
with $a_0>1$ in time domain, the scattered pulse will break in two separate
peaks. The frequency of the emitted radiation around the maxima can be
estimated as $\omega=1/(1+a^2)\approx 1/2$. The higher the pulse intensity, the
larger is the separation of the peaks. In the limit of a high instantaneous value
of $a\gg 1$, the damping of vector potential is proportional to $a^{-1}$.
Moreover, the time stretching is proportional to $a^2$. That is, the electric
field of the scattered radiation will scale as $a^{-3}$, and the corresponding
spectral intensity (see below) drops with $a$. In the limit of low $a\ll 1$, the
intensity of the scattered radiation just scales as $a^2$. Therefore, in the
case of the laser pulses with high amplitude $a_0$, we expect the spectral
intensity of the scattered light to have a pronounced maximum at relatively
high frequency arising from the scattering at the front and tail of the pulse.

\subsubsection{Spectral intensity of the back-scattered light}
\label{ssubsec:spectrum}

For the non-negative frequencies the Fourier transform of the scattered vector
potential is given by
\begin{equation}\label{eq:fourier}
    A_x(\omega) = \frac{1}{2R}\int\limits_{-\infty}^{\infty}a(\varphi)
    e^{i\omega\zeta-i\varphi} d\varphi.
\end{equation}
For the frequencies in the range $\frac{1}{1+a_0^2}<\omega<1$, the exponential
factor in (\ref{eq:fourier}) has two stationary phase
points\cite{Narozhnyi:JETP1996, Seipt:LasPhys2013}. From now on, symmetric
laser pulses centered in the origin are considered to simplify formulas, though
the same can be done for asymmetric pulses as well. Using the stationary phase
approximation, the integral in the r.h.s. of Eq.~(\ref{eq:fourier}) can be
represented in the following way:
\begin{equation}
    A_x(\omega)\approx \frac{1}{R}\sqrt{\frac{2\pi}{\omega\left| (\ln a(\varphi_\omega)^2)' \right|}}
    \cos\left(\omega\int_0^{\varphi_\omega} (a(\xi)^2)'\xi d\xi+\frac{\pi}{4}\right),
\end{equation}
where the stationary phase point $\varphi_\omega$ is defined by the following condition
\begin{equation}
    \omega = \frac{1}{1+a(\varphi_\omega)^2},
\end{equation}
\begin{equation}
    \varphi_\omega>0.
\end{equation}
The divergence of the spectrum in the low-frequency edge is a result of the
degeneracy of the corresponding stationary phase point. Much better agreement in
the low-frequency tail can be obtained in terms of the Airy
function\cite{Seipt:LasPhys2013, Narozhnyi:JETP1996}, but this is beyond the
scope of this work. The spectral intensity is then given by\cite{LandauVol2,
jackson1962classical}
\begin{equation}
    \left.\frac{d^2I}{d\omega d\Omega}\right|_{\mathrm{on-axis}}=\frac{\omega^2
    R^2}{\pi^2}|A_x(\omega)|^2\,,
\end{equation}
where $d\Omega$ stands for solid angle element. In writing this expression, both
$x$ and $y$ components were taken into account as they provide exactly the
same spectral intensity.

To extract the scaling of the spectrum with peak value of field and duration it
is convenient to introduce the normalized pulse intensity profile $f$ and pulse
duration $\tau$:
\begin{equation}
    \label{eq:af}
    a(\varphi)^2 = a_0^2f(\varphi/\tau),
\end{equation}
\begin{equation}
    f(0)=1.
\end{equation}
The expression for the intensity distribution transforms to
\begin{equation}\label{eq:dI}
    \left.\frac{d^2I}{d\omega d\Omega}\right|_{\mathrm{on-axis}}=
    \frac{2\omega\tau}{\pi}\left|y\frac{df^{-1}(y)}{dy}\right|
    \cos^2\chi\mathrm{,}
\end{equation}
where
\begin{equation}
    \chi =\omega\tau a_0^2\int\limits_y^1f^{-1}(\xi)d\xi-\frac{\pi}{4},
\end{equation}
\begin{eqnarray}
    y = \frac{1-\omega}{a_0^2\omega},\\
    0 < y < 1.
\end{eqnarray}
Here, the non-negative inverse function $f^{-1}$ of the normalized pulse intensity
profile $f$ is taken. In Eq.~(\ref{eq:dI}), the squared cosine factor defines
the interference structure in the nonlinear TS spectrum, and
the prefactor defines the envelope of the spectrum. The cosine argument can be
interpreted as the optical path difference for the light emitted at the pulse
front and at the pulse tail at the same intensity level. The number of
interference fringes can be estimated as  $\tau
a_0^2/\pi$\cite{Seipt:LasPhys2013}. Here, one sees that the pulse duration
does not affect the average shape of the spectrum, but only the overall scaling
and the frequency of the interference fringes. One can also notice, that due to
the factor $df^{-1}/dy$, a steep envelope of the incident pulse will provide
less energy in the back-scattered pulse in comparison with a gradual one. 

\subsection{Linear polarization}
\label{subsec:linear}

One can apply the machinery developed in the previous subsection also to linear
polarization of the incident pulse with the vector potential in the following
form
\begin{align}
    &A^L_x(\varphi)=a(\varphi)\cos\varphi.\\
    &A^L_y(\varphi)=0.
\end{align}
Equation (\ref{eq:At}) for the vector potential of the scattered light then can be rewritten as
\begin{equation}
    \label{eq:linAt}
    A_x(\zeta)=\frac{1}{R}\frac{a(\varphi(\zeta))\cos(\varphi(\zeta))}{1+a(\varphi(\zeta))^2\cos(\varphi(\zeta))^2}\,.
\end{equation}
\begin{equation}\label{eq:linzeta}
    \zeta=\varphi+\int\limits_0^{\varphi}a(\xi)^2\cos(\xi)^2d\xi.
\end{equation}
Here, the detector time $\zeta$ has rapidly oscillating components as a function
of the laser phase $\varphi$. This is because for linear polarization the
electron oscillates longitudinally along the $z$-axis. The retardation between
laser phase and detector time also oscillates, and the oscillations cause the
appearance of the high order harmonics in the scattered spectrum (see Fig
\ref{fig:dilation1}).

The Fourier transform of Eq.~(\ref{eq:linAt}) yields
\begin{equation}
    A_x(\omega)=\frac{1}{R}\int\limits_{-\infty}^{+\infty}e^{i\omega\zeta}a(\varphi)\cos\varphi
    d\varphi\,.
\end{equation}
Note that unlike Eq. (\ref{eq:fourier}) one has to keep the cosine without passing
to a complex amplitude in this expression because of the oscillatory part in the
exponential.  Now, one may want to eliminate the oscillations in the exponent,
arising due to the cosine term in the expression (\ref{eq:linzeta}) for $\zeta$.
Assuming, that the laser pulse amplitude is slow-varying, one can write
\begin{equation}
    \zeta\approx\varphi+\frac{1}{4}a(\varphi)^2\sin 2\varphi + \frac{1}{2}\int\limits_{0}^{\varphi}a(\xi)^2d\xi.
\end{equation}
As was done in the previous section, one can introduce the normalized intensity
profile (\ref{eq:af}) and perform the integration using the stationary phase
approximation. Doing so results in
\begin{equation}
    \label{eq:Aw_lin}
    A_x(\omega)\approx\sum\limits_{m=2k+1}(-1)^{\frac{m-1}{2}}\sqrt{\frac{4\pi\tau}{R^2\omega}
    \left|y\frac{df^{-1}}{dy}\right|_{y_m}}
    P_m\left(\frac{m-\omega}{2}\right)
    \cos\chi_m.
\end{equation}
\begin{equation}
    P_m(x) = J_{\frac{m-1}{2}}(x)-J_{\frac{m+1}{2}}(x),
\end{equation}
\begin{equation}
    \chi_m=\frac{\omega\tau a_0^2}{2}\int\limits_{y_m}^{1}f^{-1}(\xi)d\xi-\frac{\pi}{4},
\end{equation}
\begin{equation}
    y_m=2\frac{m-\omega}{a_0^2\omega}.
\end{equation}
Here $J_m$ is the Bessel function of $m$-th order.  Note, that for the
calculation of the contribution of $m$-th harmonic, the stationary phase
approximation requires $\omega\in\left(\frac{m}{1+a_0^2/2},m\right)$. From here
it follows, that the harmonics start overlapping in the spectral domain for
$m\ge 2/a_0^2$.

For the non-interfering part of the spectrum one can write
\begin{equation}
    \left.\frac{d^2I_m}{d\omega d\Omega}\right|_{\mathrm{on-axis}}=\frac{2\omega\tau}{\pi}\left|y_m\frac{df^{-1}(y_m)}{dy_m}\right|
    P_m^2\left(\frac{m-\omega}{2}\right)\cos^2\chi_m,
\end{equation}
where $m$ is an odd number. One can see, that the shape of each particular harmonic
is similar to the case of a circularly polarized pulse.  The only difference is
the factor $P_m$ containing the Bessel functions. The argument of Bessel
functions ranges from $\frac{ma_0^2}{2+a_0^2}$ to $0$ with increasing
frequency. Knowing that $J_\alpha(x)\sim x^\alpha$, one can conclude, that this
factor will lead to the suppression of high frequencies inside every harmonic
except the first one. Decrease of the maximum value of $J_\alpha(x)$ with
$\alpha$ provides the suppression of high-order harmonic emission in general.

In this section we have analyzed the time-dependence of the scattered vector
potential in the nonlinear TS.  The complete analytical expressions for the
spectral intensity of the back-scattered radiation have been written in the
stationary phase approximation. Now we turn to the numerical evaluation of the
spectra and comparison between exact and approximate solutions.

\section{Numerical calculations}
\label{sec:numerical}
\subsection{Method}
\label{subsec:method}

Since the expression for the scattered vector potential (\ref{eq:At}) is exact,
there is essentially no need to numerically solve the equations of motion for
calculation of the back-scattered spectrum. This allows us to greatly
simplify the entire calculation process. The routine for calculating the
back-scattered spectrum for any given incident laser pulse is the following: 
\begin{figure}
    \subfloat[Circular polarization]{
        \includegraphics[width=.5\linewidth]{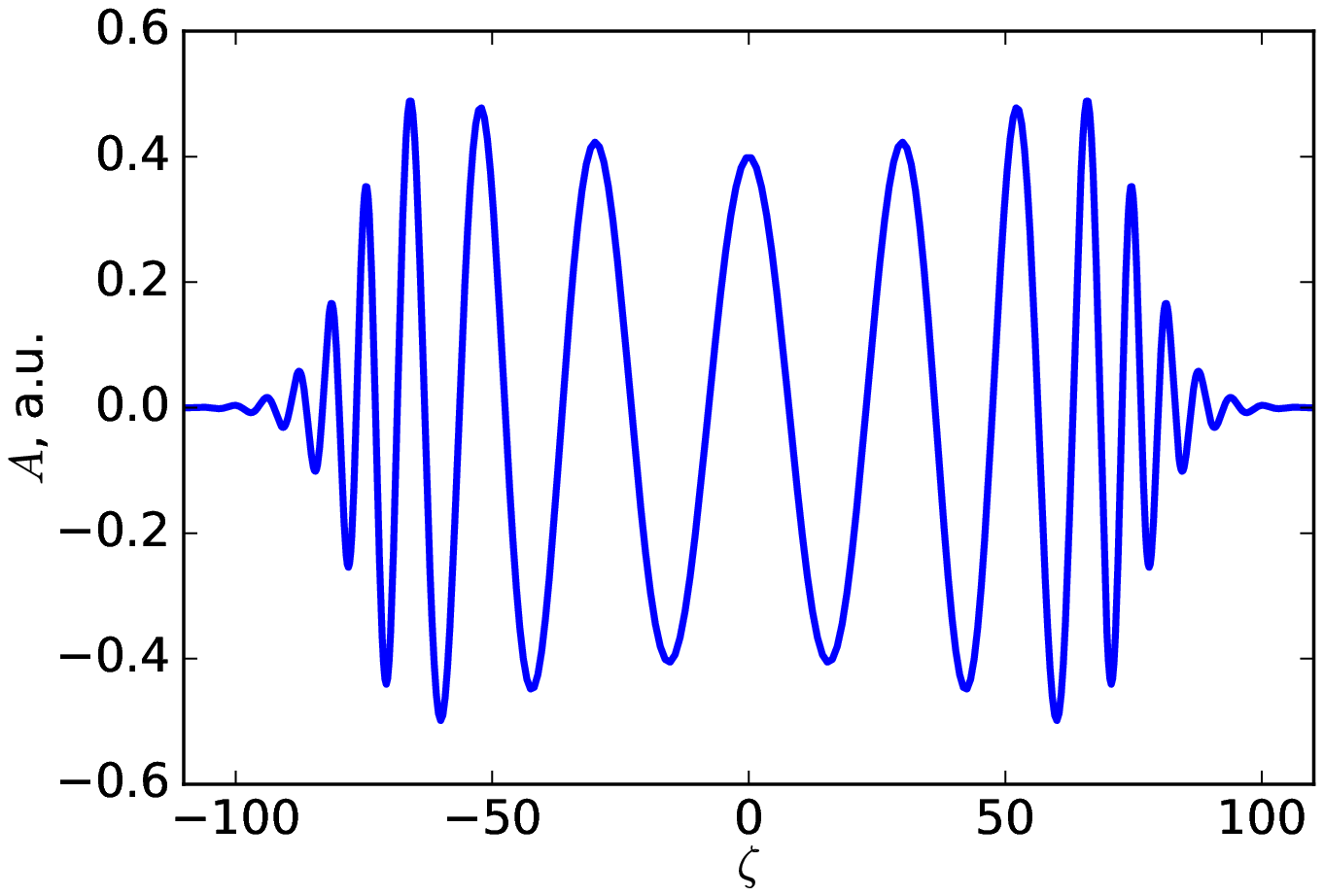}
        \label{fig:stretched0}
    }
    \subfloat[Linear polarization]{
        \includegraphics[width=.5\linewidth]{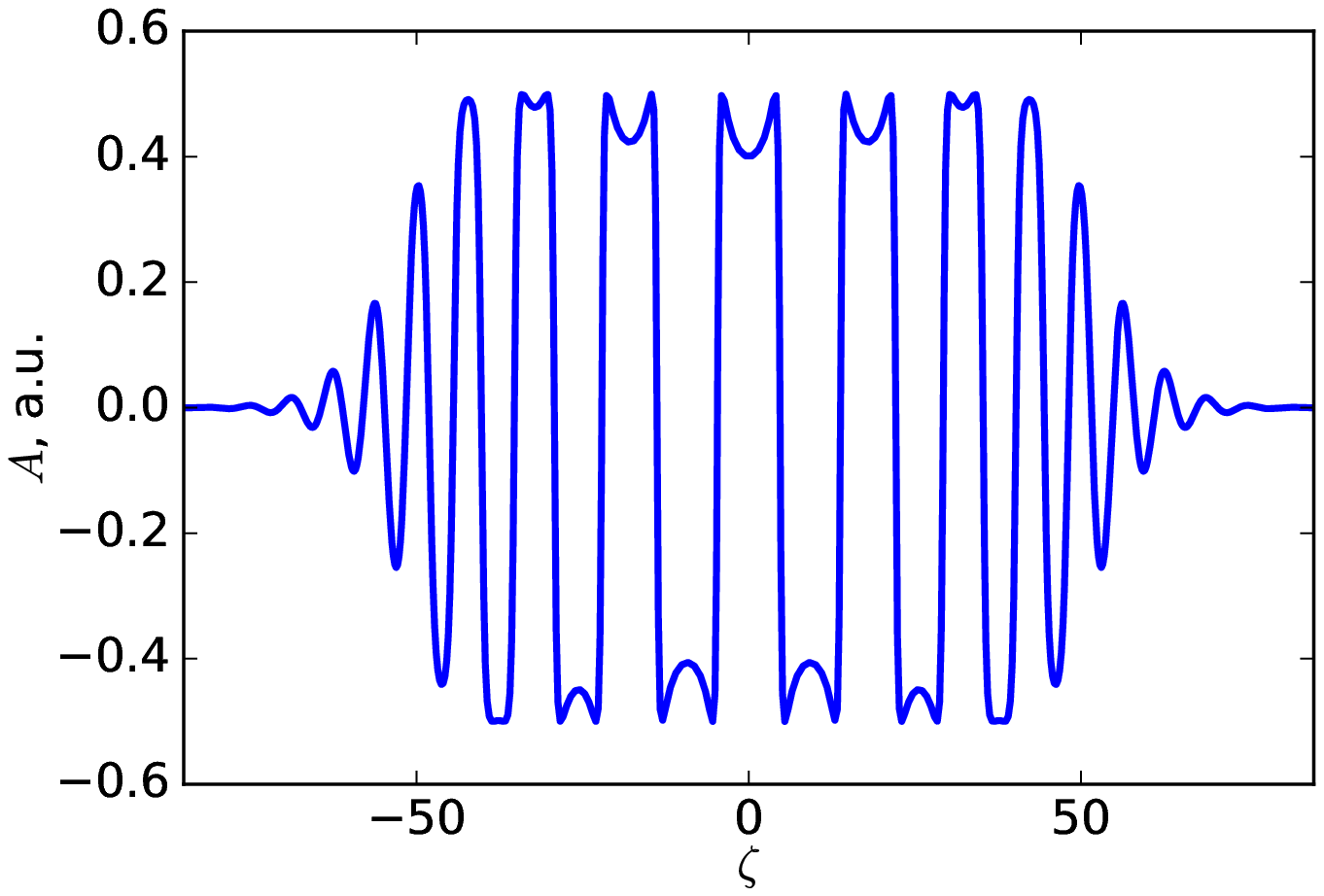}
        \label{fig:stretched1}
    }
    \centering
    \caption{Temporal structure of the scattered pulse. The incident pulse has
    Gaussian envelope, $a_0=2$, $\tau=20$. \label{fig:stretched}}
\end{figure}
\begin{itemize}
    \item Using ordinary integration routines (either numerical, or
        analytical), one obtains the $\zeta(\varphi)$ dependence. In the case
        of evenly distributed discrete points $\{\varphi_i\}$, this results in
        the discrete set of corresponding detector times
        $\{\zeta_i(\varphi_i)\}$. As discussed in Sec.  \ref{ssubsec:doppler},
        the $\{\zeta_i\}$ points will be distributed unevenly.
    \item The scattered field value at each point $\varphi_i$ is found using
        Eq.~(\ref{eq:At}). As a result, one obtains the scattered pulse
        vector-potential in the temporal domain $\{(\zeta_i,A_i)\}$
        (Figs.~\ref{fig:dilation}, \ref{fig:stretched}).
    \item Discrete dependence $\{(\zeta_i,A_i)\}$ is Fourier transformed. In
        most cases, values of $\zeta_i$ will be distributed unevenly, so to make
        this step effective one can either use the Fast Fourier Transform (FFT)
        routines in combination with interpolation, or special techniques of
        discrete Fourier transform on unevenly spaced grids. In what follows the
        first option (interpolation) was used. At this point one can notice that
        using detector time improves performance of the computations.  Namely,
        the number of operations needed to calculate on-axis spectrum using
        ordinary integration routines scales as $O(N^2)$ with $N$ being the
        number of points on the trajectory, while using detector time allows
        using FFT directly. Interpolation scales as $O(N)$, and FFT as
        $O(N\log N)$. Therefore, overall number of operations scales as $O(N\log N)$.
\end{itemize}
\subsection{Results}
\label{subsec:results}
\begin{figure}
    \subfloat[Spectral intensity of the back-scattered light.  Analytical result
    presents spectral envelope. Incident pulse has Gaussian envelope, $a_0=5$,
$\tau=200$.]{
        \includegraphics[width=.5\linewidth]{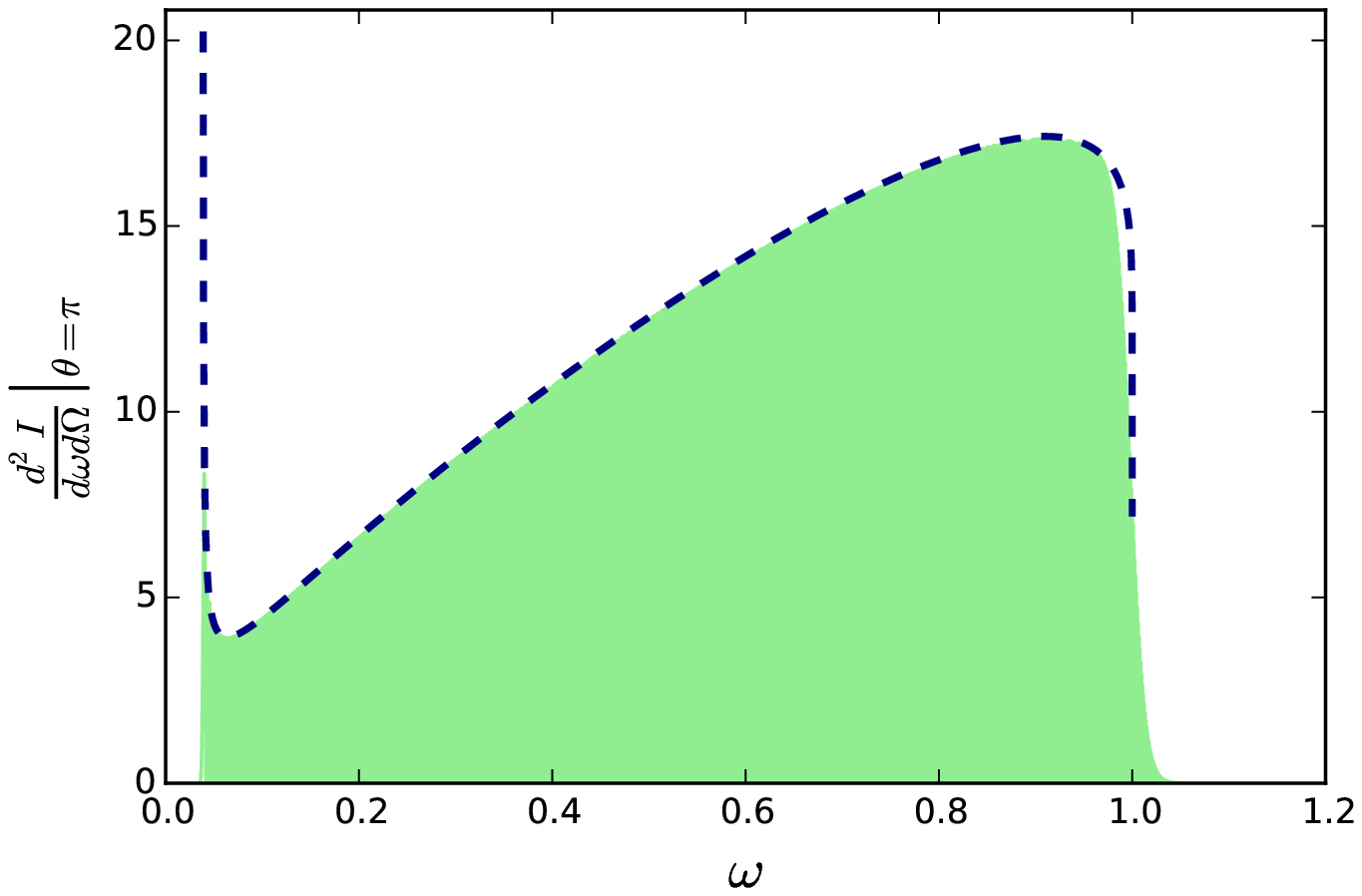}
        \label{fig:comp1}
    }
    \subfloat[Squared absolute spectrum of the vector-potential of the
            back-scattered light. Incident pulse has cosine envelope, $a_0=5$,
        $\tau=20$.]{
        \includegraphics[width=.5\linewidth]{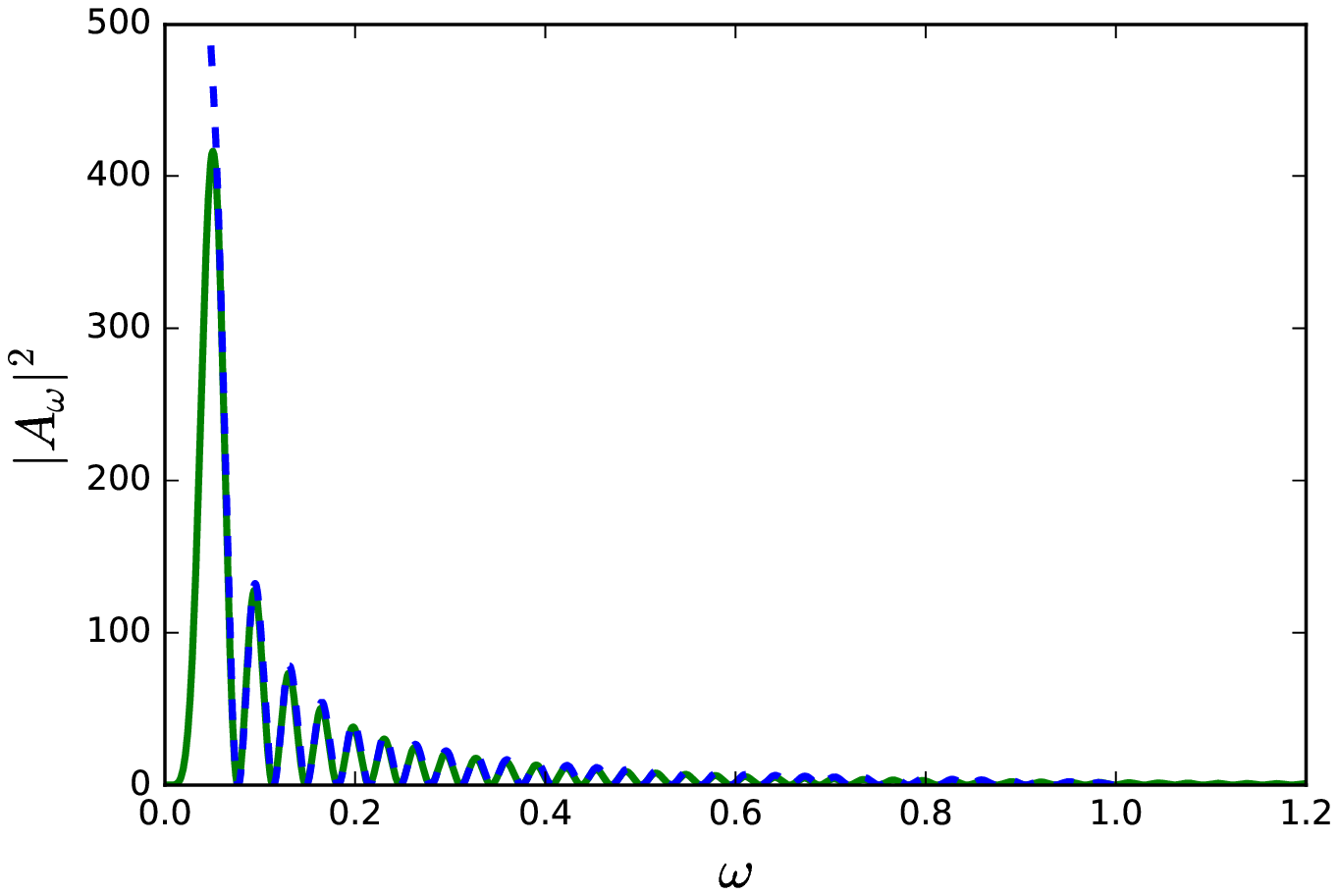}
        \label{fig:comp2}
    }
    \centering
    \label{fig:comp}
    \caption{Comparison between results of numerical simulations (solid) and stationary
    phase approximation (dashed).}
\end{figure}
From now on, we assume the case of circularly polarized light. Consider three
types of the pulse shapes, namely cosine, cosine-squared and Gaussian (see
Appendix~\ref{sec:shapes} for details). For sufficiently long pulses (with
duration $\tau\gg 1$), the stationary phase approximation yields very good
agreement with the results of the numerical calculations. It describes both the
spectral envelope (Fig.~\ref{fig:comp1}) and the interference structure
(Fig.~\ref{fig:comp2}). In the most part of described frequency region
$\omega\in(\frac{1}{1+a_0^2},1)$, the approximation is applicable. Note, that
the numerical result fills the area on Fig.~\ref{fig:comp1} due to dense
interference structure for the high values of peak intensity and duration used
in this example ($a_0=5$, $\tau=200$). Figure \ref{fig:comp2} presents the
situation where the pulse duration is lower, hence, the interference structure
is more distinguishable. There is one more notable thing about the spectrum on
Fig.~\ref{fig:comp1} - the peak of the spectral envelope at relatively high
frequencies. The origin of this peak was discussed in
Sec.~\ref{ssubsec:amplitude}.  
\begin{figure}
    \subfloat[$a_0=5$, $\tau=100$.]{
        \includegraphics[width=.5\linewidth]{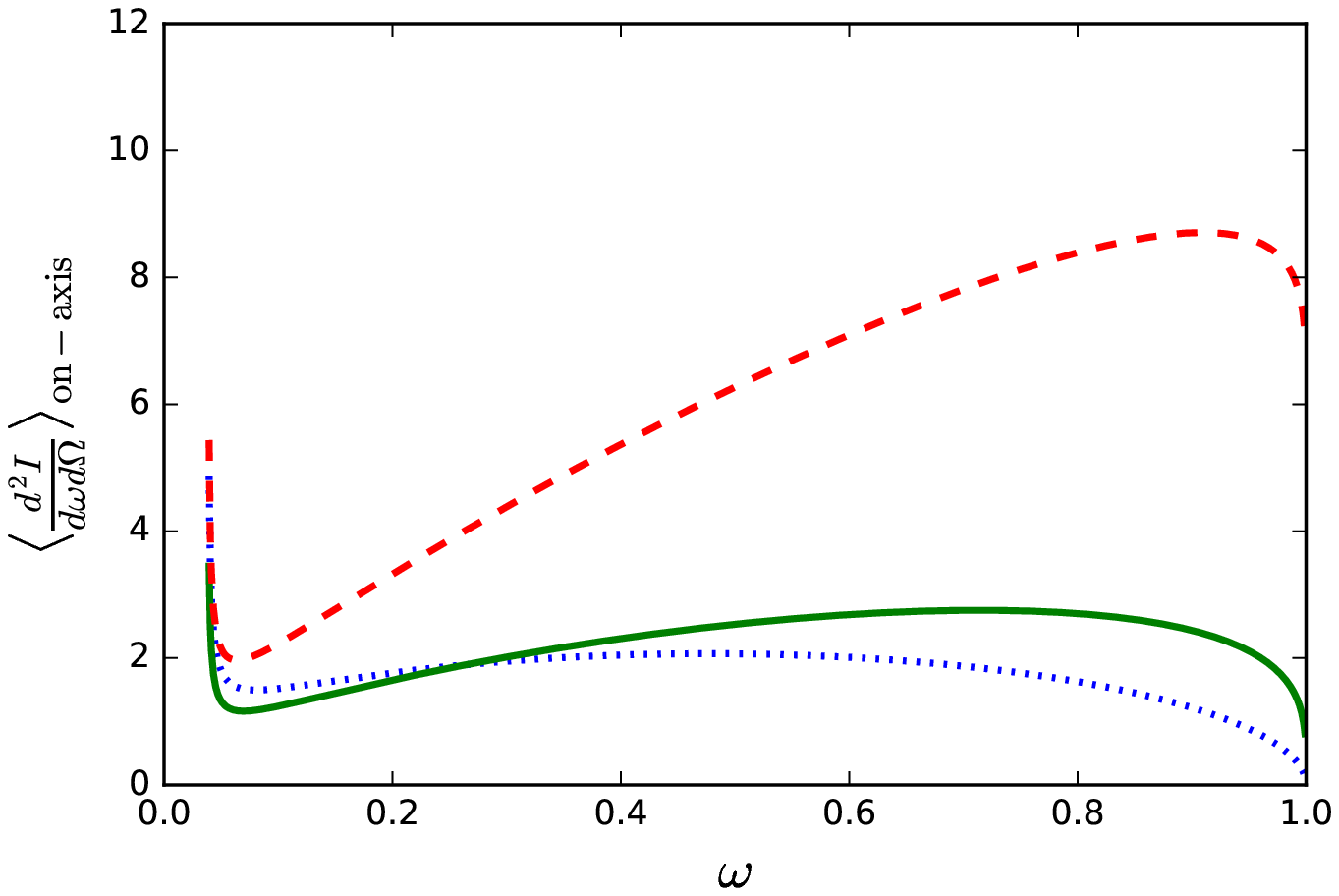}
        \label{fig:shapes_a0}
    }
    \subfloat[$\int a'(\varphi)^2d\varphi=250$,
        $\tau=50$.]{
        \includegraphics[width=.5\linewidth]{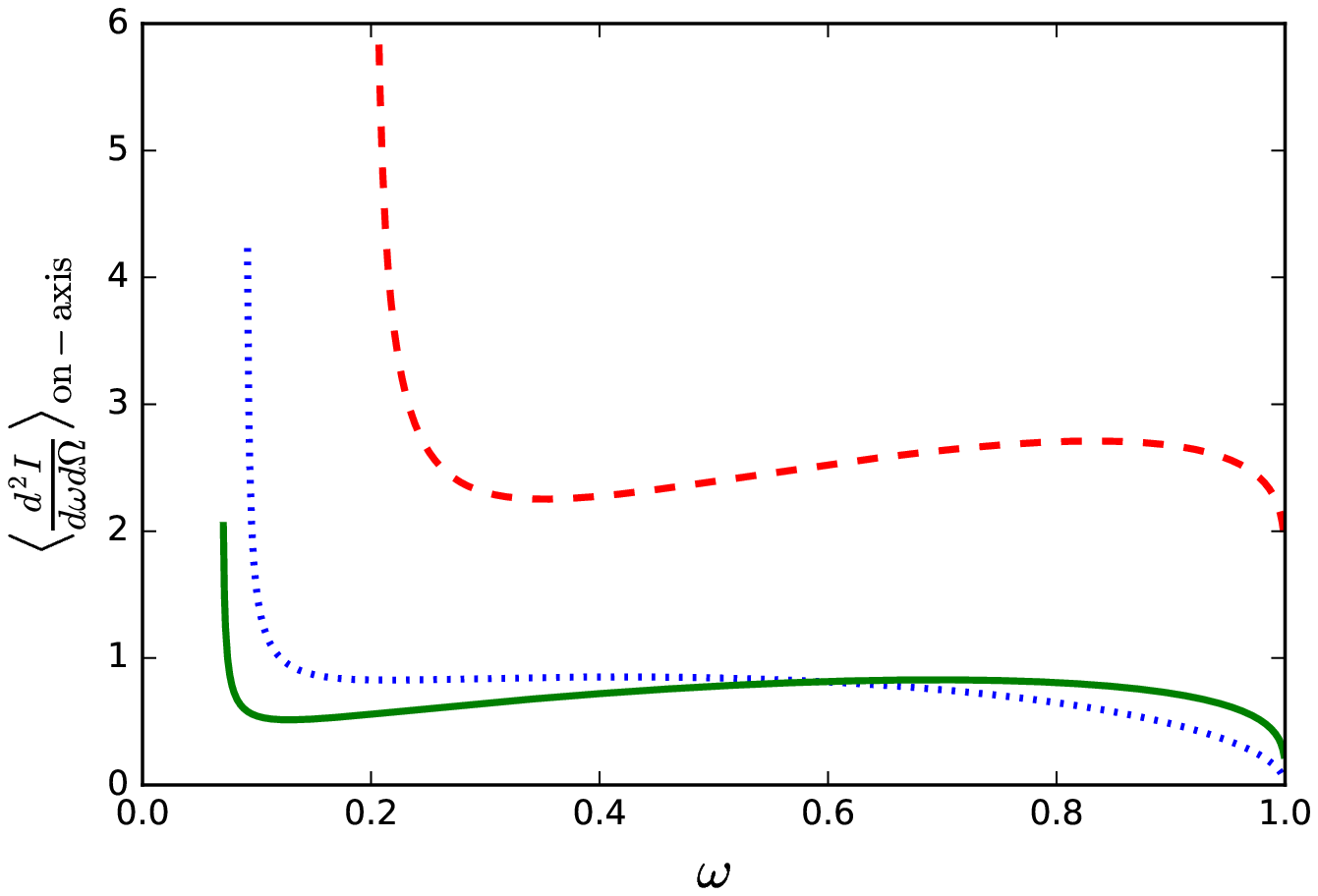}
        \label{fig:shapes_E}
    }
    \centering
    \label{fig:shapes}
    \caption{Average spectral intensity of the back-scattered radiation. 
        Fixed peak field value (a) and fixed incident pulse energy (b). 
        Incident pulse has cosine-squared (solid), Gaussian (dashed) and
        cosine (dotted) envelope. The analytic expressions are collected
        in~\ref{sec:shapes}.}
\end{figure}

The subject of great interest is the influence of the pulse shape on the
spectrum of the back-scattered radiation. Considering the incident pulse having
slowly varying amplitude, one can use equation (\ref{eq:dI}) to obtain the average
spectrum of the scattered pulse.  The average spectral intensity directly
includes the shape of the pulse. Hence, different pulse shapes with the same
peak intensity yield significantly different spectral intensity profiles
(Fig.~\ref{fig:shapes_a0}). The huge discrepancy between light scattered by
Gaussian pulse and $\cos$-pulses is due to fact that Gaussian pulse has wide,
smooth low-intensity wings in contrast to the finite support pulses. These
wings provide efficient back-scattering at relatively high frequencies, while
the low-frequency part of the spectrum is relatively suppressed. 

Assuming the incident pulse to be symmetric, one can directly reconstruct the
pulse shape from the average spectral intensity profile, up to the pulse
duration. The value of $a_0$ can be extracted from the low frequency cutoff,
and the envelope itself. Knowing $a_0$, using (\ref{eq:dI}) one can obtain the
$f^{-1}(y)$ dependence, hence the temporal laser intensity profile. One can
also see that the differences in the scattering picture occur in the case of
the fixed incident pulse energy as well (Fig.~\ref{fig:shapes_E}). This finding
can be used in high intensity laser-matter interactions experiments for laser
pulse characterization. This scheme, of course, requires circular polarization
as for linear polarization different harmonics are overlapping and the shape
of the first harmonic cannot be determined.

\section{Conclusions}
\label{sec:concl}

In this paper the nonlinear Thomson back-scattering was studied within the
classical framework.  It has been shown that interpretation of $\zeta=t+z$
variable as the detector time significantly simplifies the calculation of the
scattered radiation. It also leads to the simple way of analytical description
of the scattered spectra.  Temporal properties of the light scattered on the
single electron allow one to make qualitative conclusions about the spectrum
which stay true if the scattering occurs on many particles incoherently. Using
this approach, we compared the back-scattered spectra for various pulse
shapes, and came to the conclusion that the pulse shape strongly affects the
scattered spectrum. Under some assumptions (symmetric pulse peak, high peak
intensity and relatively high duration) pulse shape can be reconstructed from
the scattered spectral intensity, even if the interference structure is not
resolved.

We have not discussed the spectrum modification due to initial
velocity spread of the electrons, which definitely affects the back-scattered
spectrum. This effect can also be significant and can be the subject of
further investigation.

\begin{acknowledgments}
    This work was supported by the Helmholtz Association
    (Helmholtz Young Investigators group VH-NG-1037).
\end{acknowledgments}

\appendix
\section{Derivation of Eq. (\ref{eq:Aw_lin})}
Assuming the amplitude to be slowly varying one can start with the expression
\begin{equation} A_x(\omega)=\frac{1}{2R}\int\limits_{-\infty}^{+\infty} \left(
    e^{i(\omega + 1)\varphi} + e^{i(\omega-1)\varphi}\right) a(\varphi)
    \exp\left(\frac{i \omega}{4}a(\varphi)^2 \sin 2\varphi + \frac{i\omega}{2}
    \int\limits_0^{\varphi} a(\xi)^2d\xi\right)d\varphi\,.  
\end{equation}
Using the Jacobi-Anger expansion
\begin{equation}
    \exp\left(\frac{i\omega}{4}a(\varphi)^2\sin 2\varphi\right) =
    \sum\limits_{m=-\infty}^{\infty} J_m \left( \frac{\omega
    a(\varphi)^2}{4}\right) e^{2im\varphi}\,,
\end{equation}
one can rewrite this integral as a sum over harmonics:
\begin{equation}
    A_x(\omega)=\frac{1}{2R} \sum\limits_{m=-\infty}^{\infty}
    \int\limits_{-\infty}^{+\infty}\left( J_m\left(\frac{\omega a^2}{4}\right) +
    J_{m+1}\left(\frac{\omega a^2}{4}\right) \right)a(\varphi) e^{i(\omega+ 2m- 1)
        \varphi+\frac{i\omega}{2}\int a^2 d\xi}d\varphi\,.
\end{equation}
Now one can apply the stationary phase approximation and notice that
for each term the stationary point condition is $\omega=\frac{1-2m}{1+a^2/2}$:
\begin{eqnarray}
    \nonumber
    A_x(\omega) \approx \sum\limits_{m=-\infty}^{\infty} 
    \sqrt{\frac{4\pi a^2}{R^2\omega (a^2)'}}
    \left(
        J_{m}\left(\frac{\omega a^2}{2}\right) + J_{m-1}\left(\frac{\omega a^2}{4}
    \right) \right) \\
    \cos\left((\omega+2m+1)\varphi^*+\frac{\omega}{2}
    \int\limits_0^{\varphi^*} a^2d\xi -\frac{\pi}{4} \right)\,,
\end{eqnarray}
where $\varphi^*>0$ is the value obtained from the stationary phase condition.
Changing summation index to $1-2m$ and substituting $\varphi^*$ provides now Eq.
(\ref{eq:Aw_lin}).

\section{Scattered spectrum for particular pulse shapes}
\label{sec:shapes}

In the calculations of the average spectral intensity three different pulse
shapes were used. For all pulses considered, corresponding expressions can be obtained
analytically using (\ref{eq:dI}). The descriptions of pulse shapes and 
average scattered spectra for circularly polarized light are presented in Table
\ref{tb:shapes}.  
\begin{table*}
    \caption{Pulse shapes and corresponding stationary point average spectra\label{tb:shapes}}
    \begin{ruledtabular}
        \begin{tabular}{lccc}
            Name used in text & $a(\varphi)$ & $f(x)$ & $\langle\frac{d^2I}{d\omega d\Omega}\rangle_{\mathrm{on-axis}}$ \\
            \hline
            Gaussian & $a_0\exp\left(-\frac{\varphi^2}{\tau^2}\right)$ & $e^{-2x^2}$ & $\frac{\omega\tau}{4\pi}\left(\frac{1}{2}\ln\frac{a_0^2\omega}{1-\omega}\right)^{-1/2}$ \\
            cosine & $\left\{\begin{array}{ll}
            a_0\cos\frac{\pi\varphi}{\tau}, & \varphi\in\left[-\frac{\tau}{2},\frac{\tau}{2}\right] \\
            0 &\mbox{otherwise}
        \end{array}\right.$ & 
            $\left\{\begin{array}{ll}
                    \cos^2\pi x, & x\in \left[-\frac{1}{2},\frac{1}{2}\right] \\
                    0 &\mbox{otherwise}
                \end{array}\right.$ & $\frac{\omega\tau}{\pi^2}\left(\frac{a_0^2\omega}{1-\omega}-1\right)^{-1/2}$ \\
                    cosine-squared & $\left\{\begin{array}{ll}
                    a_0\cos^2\frac{\pi\varphi}{\tau}, & \varphi\in\left[-\frac{\tau}{2},\frac{\tau}{2}\right] \\
                    0 &\mbox{otherwise}
                \end{array}\right.$ & 
                    $\left\{\begin{array}{ll}
                            \cos^4\pi x, & x\in \left[-\frac{1}{2},\frac{1}{2}\right] \\
                            0 &\mbox{otherwise}
                        \end{array}\right.$ &
                            $\frac{\omega\tau}{2\pi^2}\left(\sqrt{\frac{a_0^2\omega}{1-\omega}}-1\right)^{-1/2}$
        \end{tabular}
    \end{ruledtabular}
\end{table*}

\bibliography{references}

\end{document}